\documentclass[a4paper,11pt]{article}

\usepackage[utf8]{inputenc}
\usepackage[UKenglish]{babel}
\usepackage[p,osf]{cochineal}
\usepackage[varqu,varl,var0]{inconsolata}
\usepackage[scale=.95,type1]{cabin}
\usepackage[cochineal,vvarbb]{newtxmath}
\usepackage[cal=euler,scr=rsfs]{mathalpha}

\usepackage[dvipsnames]{xcolor}
\usepackage[top=3cm, bottom=4cm, left=3cm, right=3cm]{geometry}
\usepackage{setspace}
\makeatletter
\g@addto@macro\bfseries{\boldmath}
\makeatother
\usepackage{fancyhdr}
\usepackage{amsmath,amsthm}
\usepackage{amsfonts}
\usepackage{bm}
\usepackage{slashed}
\usepackage{mathtools}
\usepackage{amsbsy}
\numberwithin{equation}{section}
\allowdisplaybreaks

\usepackage{enumitem}
\usepackage{array}

\usepackage{tabularx}

\usepackage[textfont=it]{caption} 

\usepackage{hyperref}
\hypersetup{
    colorlinks,%
    citecolor=BrickRed,%
    filecolor=Blue,%
    linkcolor=Blue,%
   	urlcolor=Blue,
   	linktoc=page
}

\usepackage{cite}

\usepackage[toc,page]{appendix}

\usepackage{graphicx}
\usepackage{float}
\usepackage{subfig}
\usepackage{tikz}
\usepackage{tikz-layers}
\usetikzlibrary{calc}
\usetikzlibrary{decorations.pathreplacing}
\usetikzlibrary{decorations.pathmorphing}
\usetikzlibrary{decorations.markings}
\usetikzlibrary{arrows.meta}
\usetikzlibrary{positioning}
\usetikzlibrary{math}
\tikzset{every picture/.style={font issue=\footnotesize},
         font issue/.style={execute at begin picture={#1\selectfont}}
        }

\usepackage{mdframed}
\usepackage[most]{tcolorbox}
	\tcbset{highlight math style={enhanced,colframe=black,colback=white,boxsep=2pt,boxrule=.5pt,sharp corners}}

\newcommand{\bfupsilon}{{\boldsymbol{\upupsilon}}}
\newcommand{\bftau}{\boldsymbol{\uptau}}
\newcommand{\hframe}{\text{\bf e}}
\newcommand{\htheta}{\boldsymbol{\uptheta}{}}
\newcommand{\bfmu}
 {\boldsymbol{\upmu}}
\newcommand{\bfomega}{\boldsymbol{\upomega}}
\newcommand{\D}{\text{d}}
\newcommand{\bfg}{{\text{\bf g}}}
\newcommand{\bfxi}{{\bm\upxi}}
 
\begin{document}

\begin{titlepage}

\begin{flushright}
{\small CPHT-RR035.042025}
\end{flushright}
\vspace{0.5cm}

\begin{center}
{\LARGE{\textbf{\textsc{ 
Carrollian-holographic Derivation of\\[4pt] Gravitational Flux-Balance Laws
}}}}
\end{center}

\vspace{7mm}
\begin{center} 
Adrien Fiorucci$^{a,}$\footnote{E-mail: \href{mailto:adrien.fiorucci@tuwien.ac.at}{adrien.fiorucci@polytechnique.edu}}, %
Simon Pekar$^{b,c,}$\footnote{E-mail: \href{mailto:simon.pekar@polytechnique.edu}{spekar@sissa.it}}, %
P. Marios Petropoulos$^{a,}$\footnote{E-mail: \href{mailto:marios.petropoulos@polytechnique.edu}{marios.petropoulos@polytechnique.edu}} %
and %
Matthieu Vilatte$^{d,}$\footnote{E-mail: \href{mailto:matthieu.vilatte@polytechnique.edu}{matthieu.vilatte@umons.ac.be}}

\normalsize
\bigskip\medskip
$^a$\textit{Centre de Physique Th\'eorique,
        \'Ecole polytechnique, \\
        Centre National de la Recherche Scientifique -- Unit\'e Mixte de Recherche 7644, \\
        Institut Polytechnique de Paris, 91120 Palaiseau Cedex, France}\\
\bigskip\medskip
$^b$\textit{International School for Advanced Studies (SISSA),\\
 Via Bonomea 265, 34136 Trieste, Italy}\\
\bigskip\medskip
$^c$\textit{Istituto Nazionale di Fisica Nucleare, Sezione di Trieste,\\
 Via Valerio 2, 34127 Trieste, Italy}\\
\bigskip\medskip
$^d$\textit{Service de Physique de l'Univers, Champs et Gravitation, Université de Mons.\\
 20 Place du Parc, 7000 Mons, Belgium}

\vspace{30pt}

\begin{abstract}\noindent
We demonstrate that the BMS evolution equations for the mass and angular momentum aspects in asymptotically flat Einstein gravity follow from local Carroll, Weyl, and diffeomorphism invariance at the null conformal boundary, upon providing a minimalistic holographic dictionary as the sole input from the bulk. This result is a significant step in the quest for a flat-space holographic correspondence and offers a geometric implementation of the radiative degrees of freedom that source the boundary theory in the presence of bulk gravitational waves. 
\end{abstract}

\end{center}

\end{titlepage}

\newpage

\setcounter{page}{2}

\begin{spacing}{0.6}

\tableofcontents

\end{spacing}

\vspace{15pt}
\noindent
\hrulefill

\setcounter{footnote}{0} 

\section{Introduction}

The holographic paradigm posits a duality between a gravitational theory and a lower-dimensional, non-gravitational field theory, suggesting that bulk quantum gravity is encoded in the dynamics of codimension-one boundary degrees of freedom. Although the correspondence remains conjectural, it has yielded striking results in the presence of a negative cosmological constant, where spacetime asymptotically approaches anti-de Sitter geometry and the dual theory is a relativistic conformal field theory. This success has motivated significant efforts over the past decade to develop \textit{flat-space holography}, \textit{i.e.}, an analogous framework for the more realistic setting of asymptotically flat spacetimes. The latter are ubiquitous gravitational models in modern physics, as they provide the kinematical arena for a broad class of phenomena, from collider physics to gravitational-wave astronomy, involving localised sources of gravity such that the spacetime geometry asymptotically approaches that of Minkowski in the far, empty region.

Implementing this construction on the foundations of the AdS/CFT correspondence encounters two main challenges. First, the codimension-one conformal boundary of flat spacetime is null, implying that the boundary theory cannot be relativistic but must instead obey the principles of \textit{Carroll physics} \cite{Leblond,SenGupta:1966qer}. The latter corresponds to the low-velocity limit of special relativity in which spacelike intervals dominate over timelike ones, effectively sending the speed of light to zero and collapsing the light-cone onto the time axis. In this context, the boundary theory is not a conformal field theory in the usual sense, but rather a Carrollian conformal field theory \cite{Duval:2014uva,Duval:2014lpa,Ciambelli:2019lap,Campoleoni:2021blr,Chen:2021xkw,Donnay:2022aba,Donnay:2022wvx,Rivera-Betancour:2022lkc,Baiguera:2022lsw,Bekaert:2022oeh,Nguyen:2023vfz,Bekaert:2024itn,Chen:2024voz}.

The second key distinction between AdS and flat-space dynamics lies in the fact that asymptotic inertial observers genuinely assess energy and momentum loss due to gravitational waves reaching null infinity. Asymptotically, the frame adapted to such observers is formed by $\bfupsilon$, the congruence of generators of null infinity and $\hframe_a$ ($a=1,2$) an orthonormal frame on the two-dimensional celestial sphere. On account of Einstein equations, one finds \cite{Trautman:1958zdi,Bondi:1962px,Sachs:1962wk,Sachs:1962zza,Tamburino:1966zz,Barnich:2010eb} (see also \cite{Ruzziconi:2020cjt,Fiorucci:2021pha,Freidel:2021fxf,Geiller:2022vto,Geiller:2024ryw} for reviews)
\begin{equation}
	\bfupsilon(M) = \tfrac14 \mathcal D_a \mathcal D_b \mathcal N^{ab} - \tfrac18 \mathcal N_{ab} \mathcal N^{ab} + \tfrac18 \mathcal D_a\mathcal D^a \mathcal R \label{eq:bondi mass loss}
\end{equation}
for the mass aspect $M$, representing the observed distribution of gravitational energy on the celestial sphere, and
\begin{equation}
\begin{split}
	\bfupsilon(N_a) &= \mathcal D_a M + \tfrac14 \mathcal C_{ab} \mathcal D^b\mathcal R - \tfrac14 \mathcal N^{bc}\mathcal D_a\mathcal C_{bc} \\
	&\quad+ \tfrac{1}{16}\mathcal D_a\big(\mathcal C_{bc}\mathcal N^{bc}\big) + \tfrac12 \mathcal D^b\big( \mathcal D_{[a}\mathcal D^c \mathcal C_{b]c} + \mathcal C^c{}_{[a}\mathcal N_{b]c} \big)
\end{split}
	\label{eq:bondi am loss}
\end{equation}
for the angular momentum aspect $N_a$. In Eqs. \eqref{eq:bondi mass loss} and \eqref{eq:bondi am loss}, $\mathcal D$ stands for the boundary Weyl-covariant derivative while $\mathcal R$ denotes the related Ricci scalar \cite{Campoleoni:2023fug}. The right-hand sides are sourced by $\mathcal C_{ab}$, the shear of outgoing null geodesics, and $\mathcal N_{ab}$ the so-called Bondi news tensor, whose precise relation to $\mathcal C_{ab}$ shall be disclosed in the main content. It encodes the strain of gravitational radiation: when non-vanishing, gravitational dynamics does not yield conservation equations on the boundary but rather \textit{flux-balance laws} for the asymptotic charges \cite{Geroch:1977big,Ashtekar:1981bq,Wald:1999wa,Barnich:2011mi}. This, once again, stands in contrast to the AdS intuition, where Einstein’s equations reduce to the covariant conservation of a boundary energy--momentum tensor. From the holographic point of view, this means that gravitational radiation would act as \textit{external sources} \cite{Troessaert:2015nia,Wieland:2020gno,Donnay:2022aba,Donnay:2022wvx,Fiorucci:talk} coupled to the boundary conformal Carrollian field theory and that are responsible for breaking the global symmetries on the boundary. This is, of course, a matter of viewpoint, and the present analysis aims at showing that the boundary symmetries are, in fact, preserved upon introducing the appropriate geometric tools to describe the boundary theory. In this letter, we shall refer to the Eqs. \eqref{eq:bondi mass loss} and \eqref{eq:bondi am loss} as \textit{Bondi--van der Burg--Metzner--Sachs (BMS) flux-balance laws} \cite{Bondi:1962px,Sachs:1962wk}.

More than the null character of the conformal boundary, the non-conservation of gravitational charges presents a significant puzzle and has led to a bifurcation in the development of flat-space holography into two seemingly disconnected approaches. The first, focused solely on radiation, has given rise to the theory of massless \textit{Carrollian amplitudes}, introduced in \cite{Donnay:2022aba,Donnay:2022wvx} and further explored in \cite{Mason:2023mti,Liu:2024nfc,Stieberger:2024shv,Ruzziconi:2024zkr,Kraus:2024gso,Liu:2024llk,Ruzziconi:2024kzo,Kraus:2025wgi,Nguyen:2025sqk}, notably via the asymptotically flat limit of AdS amplitudes \cite{Alday:2024yyj}. The complementary sector has been investigated either through group-theoretical methods \cite{Barnich:2021dta,Barnich:2022bni} or by considering the asymptotically flat limit of the AdS fluid/gravity correspondence \cite{Ciambelli:2018xat,Ciambelli:2018wre,Campoleoni:2018ltl,Mittal:2022ywl,Campoleoni:2023fug}, drawing on the substantial progress made in three-dimensional flat-space holography \cite{Li:2010dr,Barnich:2012xq,Barnich:2012rz,Barnich:2013yka,Detournay:2014fva,Bagchi:2015wna,Jiang:2017ecm,Barnich:2017jgw,Ciambelli:2020eba,Campoleoni:2022wmf,Athanasiou:2024lzr,Athanasiou:2024ykt,Petropoulos:2024jie}, where radiative sources are absent. However, aside from the heuristic attempt in \cite{Donnay:2022wvx}, there has been no proper understanding of the coupling between these sources and the remaining Carrollian degrees of freedom; consequently, no intrinsic Carrollian derivation of the key equations \eqref{eq:bondi mass loss} and \eqref{eq:bondi am loss} has been achieved. 

The aim of this work is to resolve this long-standing puzzle in flat-space holography providing such a derivation. To that end, we begin by reviewing the key geometric features of null infinity. We highlight that (part of) the boundary connection depends on extrinsic data \cite{Vogel1965,Datcourt1967,jankiewicz1954espaces,Henneaux:1979vn,Bekaert:2015xua,Hartong:2015xda,Ciambelli:2019lap}, namely $\mathcal C_{ab}$ \cite{Ashtekar:1981hw,Ashtekar:1981sf}, motivating the introduction of \textit{hypermomenta} \cite{Hehl:1976kj,Iosifidis:2019dua,Iosifidis:2020gth}, associated with variations of the effective action of these additional boundary data. This offers a natural geometric framework for implementing radiative sources and their coupling to the boundary Carrollian conformal field theory. Finally, we establish a holographic dictionary between Carrollian (hyper)momenta and bulk gravitational data \cite{Donnay:2022aba,Donnay:2022wvx,Campoleoni:2023fug} for which the Carrollian dynamical equations derived herein reproduce the BMS flux-balance laws. 

Compared to recent analyses, the novelty of our approach is threefold. First, our derivation relies purely on boundary considerations, which reveals the genuinely Carrollian origin of the Bondi flux-balance laws without employing limiting procedures in the bulk near null infinity. Second, it is developed directly in the conformally compactified spacetime, where null infinity is treated as just another null hypersurface. Finally, it introduces hypermomenta conjugate to the radiative degrees of freedom, which has the merit of clarifying and geometrically grounding their coupling to the putative holographic theory. The latter is of paramount importance for improving our understanding of the flat-space holographic correspondence. 

\paragraph*{Note.} Throughout the letter, we shall make use of transformation properties under various symmetries of geometric objects. For the sake of readability, we have collected all these technical expressions into the Appendix. The latter also provides a detailed derivation of the main result of the letter, that is Eq. \eqref{eq:flux balance general}.

\section{Geometry of null infinity}
\label{sec:Geometry of null infinity}
The conformal boundary of asymptotically flat spacetime, denoted by $\mathscr I$, is a null hypersurface, so its normal vector is also tangent to it. It therefore induces an intrinsic vector field $\bfupsilon$ on the boundary, dubbed the \textit{asymptotic field of observers}, which generates the geodesic congruence of null generators of $\mathscr I$. Consequently, the pull-back $\bfg$ of the bulk metric is degenerate in this null direction,
\begin{equation}
	\bfg (\bfupsilon,\cdot) = \bm 0. \label{eq:carroll}
\end{equation}
Moreover, since the structure on $\mathscr I$ emerges from the conformal compactification procedure \cite{PhysRevLett.10.66,Penrose:1965am}, it is endowed with a conformal structure on which Weyl rescalings by an arbitrary smooth non-vanishing function $\mathcal B$ are at work \footnote{Eq. \eqref{eq:weyl} means that $\bfg$ and $\bfupsilon$ are \textit{Weyl-covariant objects} of respective weights $-2$ and $1$.}
\begin{equation}
	\bfg\mapsto \mathcal B^{-2}\bfg,\quad \bfupsilon\mapsto \mathcal B\bfupsilon, \label{eq:weyl}
\end{equation}
mapping physically indistinguishable boundary Carroll structures. From \eqref{eq:carroll} and \eqref{eq:weyl}, null infinity is endowed with a \textit{conformal Carroll structure} \cite{Duval:2014uva,Duval:2014lpa,Ciambelli:2019lap}, which provides a natural distinction between longitudinal and transverse quantities in the following sense. Introducing a clock form $\bftau$ related to the field of observers $\bfupsilon$ such that $\bftau(\bfupsilon) = 1$ \cite{Bekaert:2015xua,Ciambelli:2019lap}, transverse vectors at any point $P\in\mathscr I$ are designed to belong to 
\begin{equation}
	\text H_P(\mathscr I) = \{\text{\bf V}\in T_P\mathscr I\,|\,\bftau(\text{\bf V}) = 0\}. \label{eq:cut}
\end{equation}
To connect with the usual nomenclature, we shall refer to the two-dimensional manifold spanned by integral curves of transverse vectors at $P$ as the \textit{cut} of $\mathscr I$ at $P$, and take it to be homeomorphic to two-spheres. We shall denote by $\{\hframe_a\}$ for $a=1,2$ an orthonormal frame on $\text H_P(\mathscr I)$, in the sense that the boundary metric expands as $\bfg = \delta_{ab}\htheta^a\otimes\htheta^b$ in the dual co-frame $\{\htheta^a\}$, \textit{i.e.} such that $\htheta^a(\bfupsilon) = 0$ and $\htheta^a(\hframe_b) = \delta^a{}_b$. 

We refer to $\{\hframe_A\} = \{\bfupsilon,\hframe_1,\hframe_2\}$ as a local \textit{Carroll--Cartan frame} at the point $P$ of the boundary ($A = 0,1,2$) and by definition its dual co-frame is given by $\{\htheta^A\} = \{\bftau,\htheta^1,\htheta^2\}$. In spite of the metric being degenerate, the boundary volume form is well-defined and reads $\bfmu = \bftau\wedge\htheta^1\wedge\htheta^2$ in terms of it. There are several local transformations at work on the defined boundary local frame. First, one can rotate the transverse basis independently of the longitudinal components
\begin{equation}
	\delta_r\bfupsilon = \bm 0 = \delta_r\bftau,\quad \delta_r\hframe_a = r_a{}^b \hframe_b,\quad \delta_r\htheta^a = r^a{}_b\htheta^b , \label{eq:rotation}
\end{equation}
where $r_{ab}\in\mathfrak{so}(2)$, $r_{(ab)} = 0$. Second, the clock form is genuinely defined up to a transverse one-form $\lambda_a\htheta^a$, which implies the existence of the following transformation:
\begin{equation}
	\delta_\lambda\bfupsilon = \bm 0 = \delta_\lambda \htheta^a,\quad \delta_\lambda\hframe_a = \lambda_a\bfupsilon,\quad \delta_\lambda\bftau = - \lambda_a\htheta^a. \label{eq:boost}
\end{equation}
The transformations \eqref{eq:rotation} and \eqref{eq:boost} correspond to the action of the homogeneous Carroll algebra $\mathfrak{carr}(3)\simeq \mathfrak{iso}(2)$ on the frame. In particular, $\lambda_a$ parameterises a local Carroll boost. The existence of Weyl transformations \eqref{eq:weyl} on the boundary also affects the local Carroll--Cartan basis infinitesimally as
\begin{equation}
	\delta_B \hframe_A = B \hframe_A ,\quad \delta_B \htheta^A = -B\htheta^A . \label{eq:weyl on frame}
\end{equation}
We discuss the implications of these local transformations, seen as variational symmetries of the boundary theory, in the next section.

\begin{description}
	\item[Remark.] Contrary to their timelike and spacelike counterparts, the normal vector to a null hypersurface cannot be canonically normalised. Therefore, as the normal vector is also tangent in the null case, this unavoidable ambiguity induces another local transformation
\begin{equation}
	\delta_\eta\bfupsilon = \eta\bfupsilon,\quad \delta_\eta \bftau = -\eta\bftau,\quad \delta_\eta\hframe_a = \bm 0 = \delta_\eta \htheta^a.
\end{equation}
However, because of its extrinsic status compared to the previous local transformations, it shall not be considered as a symmetry of the boundary theory in what follows. \hfill $\square$
\end{description}
The boundary basis obeys the following structure equations
\begin{equation}
	[\bfupsilon,\hframe_a] \equiv \varphi_a\bfupsilon - c^b{}_a \hframe_b,\quad [\hframe_a,\hframe_b] \equiv 2\varpi_{ab}\bfupsilon + c^c{}_{ab}\hframe_c, \label{eq:non holonomy}
\end{equation}
where $\varpi_{ab}$ measures the non-integrability of the distribution of cuts \eqref{eq:cut} on the flow of $\bfupsilon$ (whenever $\varpi_{ab}\neq 0$, the cuts do not define a spacelike foliation of $\mathscr I$). The evolution of the transverse geometry under the geodesic flow generated by $\bfupsilon$ is encoded into the rank-two tensor
\begin{equation}
	c_{ab} = \tfrac12 \theta \delta_{ab} + \xi_{ab} + c_{[ab]}, \label{eq:def cab}
\end{equation}
where $\theta$ measures the expansion of the null congruence $\bfupsilon$ and the transverse trace-free tensor $\xi_{ab}$ its shear. On account of Einstein's equations,\footnote{So far, we have only imposed the Einstein equations that provide, in Ashtekar's terms~\cite{Ashtekar:1981hw,Ashtekar:1981bq}, the ``kinematic arena'' for formulating the holographic correspondence, \textit{i.e.}, the nature of the boundary (conformal Carroll) and the connection induced on the conformal boundary by the bulk physical Levi--Civita connection. These data must be specified prior to formulating any holographic theory that is dual to Einstein gravity. In AdS, for instance, one typically imposes, consistently with the Einstein equations, that the conformal boundary is Lorentzian, with an effective speed of light proportional to the cosmological constant, and that the boundary connection is the Weyl--Levi--Civita connection.} the latter vanishes identically on the boundary \cite{Geroch:1977big,Ashtekar:1981hw,Penrose_Rindler_1986}. Therefore, the bulk Levi--Civita connection naturally induces an affine connection $\nabla$ on $\mathscr I$ which is torsion-free and obeys \cite{Ashtekar:1981hw,Ashtekar:1981sf}
\begin{equation}
	\nabla \bfg - 2 \bm\upalpha\otimes \bfg =\bm 0,\quad \nabla \bfupsilon + \bm\upalpha\otimes \bfupsilon = \bm 0. \label{eq:strong weyl}
\end{equation}
The boundary connection is therefore both metric- and field-of-observer-compatible \textit{up to Weyl rescalings} \cite{Weyl:1918ib,10.4310/jdg/1214429379,10.1063/1.529582}. For the definition \eqref{eq:strong weyl} to be compatible with the boundary Weyl structure, the boundary one-form $\bm\upalpha$ transforms as \mbox{$\bm\upalpha \mapsto \bm\upalpha - \D \ln\mathcal B$} under Weyl rescaling \eqref{eq:weyl}. The boundary connection $\nabla$ is therefore naturally a \textit{Weyl-affine Carroll connection} \cite{Ciambelli:2018wre}. Einstein equations further impose that $\alpha_0 = \tfrac12 \theta$ while the spacelike Weyl connection $\alpha_a$ is encoded into purely bulk metric data and thus constitutes another boundary background field that contributes to the definition of $\nabla$. In the following, we shall assume that $\bm\upalpha$ is closed for simplicity.

Now comes a crucial observation: if $\bfomega^A{}_B$ denote the connection one-forms of $\nabla$ in the Carroll--Cartan frame, the vanishing of the torsion and the conditions \eqref{eq:strong weyl} are unsufficient to constrain completely $\beta_{ab} \equiv \bfomega^0{}_b(\hframe_a)$, as they leave $\beta_{(ab)}$ arbitrary. In Carroll geometry, this property is understood as the statement that torsion-free compatible Carroll connections constitute an affine space modelled on transverse symmetric rank-two tensors \cite{Vogel1965,Datcourt1967,jankiewicz1954espaces,Henneaux:1979vn,Bekaert:2015xua,Hartong:2015xda,Ciambelli:2019lap}. The transverse part of $\bfomega^0{}_a$ is therefore \textit{not} entirely fixed from purely boundary considerations. However, relating the boundary conformal manifold $\mathscr I$ to a bulk geometry allows to resolve this apparent ambiguity and give a physical interpretation of it. Indeed, it can be proven that $\beta_{ab}$ coincides with the boundary value of the deviation tensor of the congruence of outgoing null geodesics generated by a vector field $\text{\bf k}$, such that the pull-back of the associated one-form $\text{\bf k}_\flat$ to $\mathscr I$ is equal to (minus) the clock form $\bftau$. This observation derives from an explicit computation of the induced connection on $\mathscr I$ from the bulk Weyl--Levi--Civita connection on the unphysical conformally compactified spacetime using the so-called \textit{rigging method} introduced in \cite{Mars:1993mj} (see also \cite{Chandrasekaran:2018aop,Chandrasekaran:2021hxc,Freidel:2022vjq,Ciambelli:2023mir}), the rigging vector being $\text{\bf k}$. In particular, its traceless part, denoted as $\beta_{\langle ab \rangle}$, is the \textit{Bondi asymptotic shear} and encodes the two degrees of freedom of bulk gravitational radiation \cite{Ashtekar:1981hw}. This concludes our presentation of the relevant kinematical aspects of the boundary geometry.

\section{Carrollian momenta and dynamics}
\label{sec:dynamics}
We now turn to the derivation of the dynamical equations that the dual conformal Carrollian field theory is committed to obey. Let $S$ be an action for this theory coupled to the background geometry. In the perspective of \cite{Ciambelli:2018xat,Ciambelli:2018wre,Campoleoni:2018ltl,Ciambelli:2020eba,Petkou:2022bmz,Campoleoni:2022wmf,Mittal:2022ywl,Campoleoni:2023fug}, we are agnostic of the microscopic ``matter'' field content of this theory, which we denote collectively by $\Phi$. However, we demand that it is simultaneously invariant under local Carroll transformations \eqref{eq:rotation} and \eqref{eq:boost}, Weyl transformations \eqref{eq:weyl on frame} and diffeomorphisms to extract general features on the boundary theory. 

Taking an arbitrary variation of $S$ yields
\begin{equation}
	\delta S = \frac{1}{16\pi G}\int_{\text{$\mathscr I$}}\bfmu \left(\mathcal E_\Phi\delta\Phi + \delta\htheta^A(\text{\bf T}_A) + \delta\bfomega^A{}_B(\bm\Omega_A{}^B) \right) \label{eq:delta S}
\end{equation}
where $G$ is the Newton--Cavendish constant in four dimensions. The first term in Eq. \eqref{eq:delta S}. represents the contribution of the ``matter'' fields and $\mathcal E_\Phi \approx 0$ are their equations of motion. From now on, we shall denote by $\approx$ equalities that hold on-shell for them. Next, variations with respect to the boundary background geometry, which is fully encoded in the Carroll--Cartan co-frame, define the \textit{Carrollian momenta} \cite{Ciambelli:2018wre,Ciambelli:2018xat,Ciambelli:2018ojf,Petkou:2022bmz} that we expand as
\begin{equation}
	\text{\bf T}_0 = \Pi\bfupsilon + \Pi^a\hframe_a,\quad \text{\bf T}_a = P_a\bfupsilon + \Pi^b{}_a\hframe_b.
\end{equation}
They respectively stand for the energy density ($\Pi$), the energy flux ($\Pi^a$), the momentum density ($P_a$) and the stress tensor ($\Pi^a{}_b$). Finally,
we profit from the main lesson of the previous section, which was that (part of) the boundary connection as an independent field with respect to the geometry. This is why we introduce the vector fields $\bm\Omega_A{}^B$ as the response to fluctuations of the boundary connection one-forms $\bfomega^A{}_B$, and we refer to them as the \textit{hypermomentum} \cite{Hehl:1976kj,Iosifidis:2019dua,Iosifidis:2020gth}. For $S$ to be Weyl invariant, we shall assume that 
\begin{equation}
	\text{\bf T}_A\mapsto \mathcal B^4 \text{\bf T}_A,\quad \bm\Omega_A{}^B \mapsto \mathcal B^3\bm\Omega_A{}^B,
\end{equation}
\textit{i.e.}, that the Carrollian momenta and hypermomenta transform as weight-four and three Weyl-covariant objects. At this stage, it is important to emphasise that the dynamical fields and the geometry play different roles in the variational principle, Eq. \eqref{eq:delta S}. Specifically, both the frame $\htheta^A$ and the connection $\bfomega^A{}_B$ are treated as background fields, since, unlike $\Phi$, they are not subject to equations of motion. Nonetheless, their variations are retained in the variational principle, as they are required to define the associated momenta and derive the constraints dictated by local symmetries (\textit{Noether identities}).\footnote{Note that this is analogous to the standard treatment of matter fields coupled to a background Lorentzian geometry, where the covariant (or Einstein--Hilbert) energy--momentum tensor is obtained by varying the matter action with respect to the background metric. Its conservation then follows from diffeomorphism invariance and the matter equations of motion. The only difference here is that the connection includes components not determined by the geometry, thereby increasing the number of background fields.}

This is precisely what we now set out to do, \textit{i.e.} we study the constraints imposed on the Carrollian momenta by the local symmetries. Let us start with local Carroll boost invariance. Contracting Eq. \eqref{eq:delta S} with the transformations \eqref{eq:boost} for the frame and \eqref{eq:omega boost} for the connection and performing an integration by parts to isolate the boost parameter, we obtain
\begin{equation}
	\delta_\lambda S = \mathscr N_\lambda[\mathcal E_\Phi] - \frac{1}{16\pi G} \int_{\mathscr I}\bfmu\,\lambda_a\, \left( \Pi^a + \pmb{\mathcal D}\cdot\bm\Omega_0{}^a\right),
\end{equation}
where $\mathscr N_\lambda[\mathcal E_\Phi]$ is a linear functional in the equation of motion for the fields $\Phi$ and $\pmb{\mathcal D}$ stands for the Weyl-covariant derivative on the boundary. It acts as
\begin{equation}
	\pmb{\mathcal D} = \bm\nabla + w\bm\upalpha
\end{equation}
on weight-$w$ quantities and $\pmb{\mathcal D}\cdot \text{\bf V} \equiv \mathcal D_{\bfupsilon} V^0 + \mathcal D_a V^a$ denotes the divergence of any $\text{\bf V} \in T\mathscr I$. Because $\lambda_a$ is an arbitrary function on $\mathscr I$, requiring local Carroll boost invariance yields
\begin{equation}
	\delta_\lambda S = 0\quad\Rightarrow\quad \Pi^a \approx - \pmb{\mathcal D}\cdot\bm\Omega_0{}^a, \label{eq:energy flux}
\end{equation}
when the fields $\Phi$ are on-shell and therefore, the functional $\mathscr N_\lambda$ vanishes identically. Crucially, the fact that $\Pi^a\neq 0$ is \textit{not} in contradiction with local Carroll boost invariance, which is often thought of as synonymic of the absence of energy flux \cite{deBoer:2017ing,deBoer:2021jej,deBoer:2023fnj}. Our analysis shows that the latter statement is generally incomplete and should be replaced by ``Carroll-boost symmetry implies that energy flux is of geometric nature only.'' This subtle feature, already derived in \cite{Ciambelli:2018wre,Mittal:2022ywl,Campoleoni:2023fug} from direct computations, is rooted in the fact that Carroll connections possess degrees of freedom that cannot be fixed by the background geometry, namely $\beta_{(ab)}$, as we stressed earlier.

We can apply similar techniques to derive the remaining Noether identities, namely for rotation, Weyl and diffeomorphism invariance. First, rotation and Weyl symmetries further respectively fix the skew-symmetric and trace-full parts of the stress tensor as
\begin{equation}
	\Pi_{[ab]} \approx \pmb{\mathcal D}\cdot \bm\Omega_{[ab]} , \quad \Pi + \Pi^a{}_a \approx - \pmb{\mathcal D}\cdot \big(\bm\Omega_0{}^0 + \bm\Omega_a{}^a\big),
	\label{eq:constraints rot weyl}
\end{equation}
where the spacelike indices are lowered in with $\delta_{ab}$. Furthermore, diffeomorphism invariance, \textit{i.e.} $\mathscr L_\bfxi S = 0$, gives rise to the sought-after dynamical equations \cite{Ciambelli:2018xat,Ciambelli:2018wre,Petkou:2022bmz}, which take the form of flux-balance equations for the energy--momentum densities as
\begin{equation}
	\pmb{\mathcal D}\cdot \text{\bf T}_0 \approx \pmb{\mathcal R}^A{}_B(\bfupsilon,\bm\Omega_A{}^B),\quad \pmb{\mathcal D}\cdot \text{\bf T}_a \approx \pmb{\mathcal R}^B{}_C(\hframe_a,\bm\Omega_B{}^C).
	\label{eq:flux balance general}
\end{equation}
Further details about their derivation are provided in Appendix~B. As the latter involves explicitly the constraints \eqref{eq:energy flux}--\eqref{eq:constraints rot weyl}, these equations are genuinely Carroll and Weyl covariant. The right-hand sides display the curvature two-form $\pmb{\mathcal R}^A{}_B(\cdot,\cdot)$ associated with the boundary Weyl connection.\footnote{Note that flux-balance equations sharing the same structure as Eqs. \eqref{eq:flux balance general} appear while considering the constraints implied by diffeomorphism invariance for relativistic theories coupled with more general connections, \textit{i.e.}, without requiring metric compatibility or a vanishing torsion, see \cite{Iosifidis:2019dua,Iosifidis:2020gth} and more recently \cite{Iosifidis:2025sjx}.} Already at this stage, it is worth noticing that since the boundary connection is not completely determined in terms of intrinsic data, the flux terms cannot be completely recast as a divergence.\footnote{Performing a thorough analysis, one can check that the right-hand sides of Eqs. \eqref{eq:flux balance general} is a divergence if and only if the hypermomentum $\bm\Omega_0{}^a$ associated, in particular, with the ``ambiguous'' part of the boundary connection vanishes. As we shall see below, the latter is related to the news tensor which encodes the gravitational radiation: sending it to zero would therefore strongly reduce the bulk solution space.} In other words, there exists no local modification of the Carrollian momenta that would transform \eqref{eq:flux balance general} into covariant conservation equations. This feature was observed long ago in \cite{Duval:1976ht} while studying Galilean fluid dynamics. Upon identifying the Carrollian (hyper)momenta with the appropriate gravitational data, the dynamical equations \eqref{eq:flux balance general} are nothing but the celebrated BMS evolution equations at null infinity. We prove this statement in the next section by deriving the precise form of the fluxes in the right-hand sides within a concrete boundary gauge fixing. Therefore, the equations \eqref{eq:flux balance general} should be understood as the flat-space avatar of the conservation of the holographic energy--momentum tensor in the AdS/CFT framework.

\begin{description}
	\item[Remark.] Among the connection one-forms, $\bfomega^a{}_0$ and $\bfomega_{\langle ab\rangle}$ transform homogeneously under local Carroll and Weyl transformations and can therefore be safely set to zero without breaking any of these symmetries. Consequently, these variables can always be ignored in the variational principle \eqref{eq:delta S}, and the related hypermomenta, $\bm\Omega_a{}^0$ and $\bm\Omega_{\langle ab \rangle}$, set to zero, which we assume in what follows. \hfill $\square$
\end{description}

\section{Gravitational flux-balance laws}
\label{sec:Gravitational flux-balance laws}
For definiteness, we opt for relating the Carrollian dynamical equations \eqref{eq:flux balance general} to the BMS flux-balance laws in the \textit{BMS frame} for which the clock form is exact so that both $\varphi_a$ and $\varpi_{ab}$ identically vanish, making the distribution of cuts integrable on $\mathscr I$. So, there exists a smooth function $u$ on $\mathscr I$ such that $\bftau = \D u$, which defines a coordinate along the null direction. Boundary coordinates $(u,x^i)$ are fixed upon providing a given coordinate system $(x^i)$ on the cuts. Furthermore, we require that the clock form is lifted as $\text e^{-b} \D u$ (where the function $b$ vanishes at $\mathscr I$) in the bulk, which corresponds to choosing null Gaussian normal coordinates $(r,u,x^i)$ in the vicinity of future null infinity. By construction, $r$ is a parameter along the null geodesics generated by $\text{\bf k}$, and making a choice for $r$ would complete the gauge fixing, which we do not discuss here.\footnote{Namely, if $r$ is affine, then $\beta = 0$ everywhere and our gauge conditions are those fixing Newman--Unti coordinates \cite{Newman:1962cia,Barnich:2011ty,Ciambelli:2018wre}. Furthermore, if $r$ is Sachs' luminosity distance (\textit{i.e.}, the radius of expanding null spheres), then our gauge fixing leads to Bondi--Sachs coordinates \cite{Bondi:1962px,Sachs:1962wk}.}  

For these conditions to emerge from the bulk gauge fixing, boundary diffeomorphisms $\bm\bfxi$ act on the boundary in such a way to preserve the BMS frame. This can be achieved by compensating the change induced by the Lie derivative by appropriate local boost,  rotation and Weyl transformations to put the frame back to its original value.\footnote{This is a usual mechanism in Carroll physics, see \textit{e.g.} \cite{Henneaux:2021yzg} for a detailed explanation.} The key point is that the fitting local and Weyl transformations shall explicitly depend on the diffeomorphism at hand, as we shall see below. The \textit{gauge-fixed variation} under boundary diffeomorphisms $\delta_\bfxi$ is therefore defined as the appropriate combination of diffeomorphisms and compensating local gauge transformations that fix the (co)frame, \textit{i.e.}
\begin{equation}
	\delta_\bfxi \equiv \mathscr L_\bfxi + \delta_{\lambda(\bfxi)} + \delta_{r(\bfxi)} + \delta_{B(\bfxi)} :\quad \delta_\bfxi \bftau = \bm 0 = \delta_\bfxi \htheta^a. \label{eq:gauge fixed var}
\end{equation}
Decomposing the diffeomorphism generator as $\bm\bfxi = f\bfupsilon + Y^a\hframe_a$ in the Carroll--Cartan frame, we get that in order for the gauge-fixed variation to indeed define a gauge-fixing, the local parameters must be determined in terms of the components of the diffeomorphism generator as
\begin{equation}
	\lambda_a(\bm\bfxi) = \hframe_a(f),\ \ B(\bm\bfxi) = \bfupsilon(f), \ \ r_{ab}(\bm\bfxi) = \delta_{c[b}\htheta^{c}(\mathscr L_{\hframe_{a]}}\bm\bfxi), 
\end{equation}
and that the diffeomorphism generator must also satisfy the following constraints
\begin{equation}
	\htheta^a(\mathscr L_\bfupsilon\bm\bfxi) = 0,\quad \delta_{c(a} \htheta^{c}(\mathscr L_{\hframe_{b)}}\bm\bfxi) = \delta_{ab} B(\bfxi). \label{eq:constraints on parameters}
\end{equation}
The first one implies that the spatial components of the diffeomorphism parameter are invariant along the longitudinal direction; $\bfxi$ therefore generates a Carrollian diffeomorphism in the sense of \cite{Ciambelli:2018xat,Ciambelli:2019lap}. The second one can be rewritten as $\mathscr L_\bfxi \bfg = 2 B(\bfxi) \bfg$, implying that $Y^a$ solves the conformal Killing equation on each cut of $\mathscr I$.

For simplicity and to align with the seminal references \cite{Geroch:1977big,Ashtekar:1981hw,Penrose_Rindler_1986} we further set $\theta = 0$ from now on at the price of restricting the Weyl freedom at the boundary to functions that are invariant along the null direction, $\bfupsilon(B) = 0$. We also choose to set $c_{[ab]}=0$ which restricts rotation transformations to $\bfupsilon (r_{ab}) = 0$. Both conditions are compatible with the previous equations defining the gauge-fixed variation. The boundary connection one-forms, solution of Eq. \eqref{eq:strong weyl}, expand as
\begin{subequations} \label{eq:boundary co}
	\begin{align}
		\bfomega^0{}_0 &= -\alpha_a\htheta^a, & \bfomega^0{}_a &= -\alpha_a\bftau -\tfrac12 \mathcal C_{ab}\htheta^b, \\
		\bfomega^a{}_0 &= \bm 0, & \bfomega^a{}_b &= \gamma{}^a{}_{cb}\htheta^b ,
	\end{align}
\end{subequations}
where $\mathcal C_{ab}$ is symmetric and taken trace-free by choice of a subleading Weyl rescaling \footnote{By this, we mean a transformation $\text{\bf k}\mapsto \text e^W \text{\bf k}$ where the function $W$ vanishes at the boundary.} in the direction of $\text{\bf k}$, see, \textit{e.g.}, \cite{Barnich:2011ty,Barnich:2012nkq}. As mentioned above, it corresponds to the Bondi shear of the dual asymptotically flat spacetime. The purely horizontal connection coefficients
\begin{equation}
	\gamma{}^a{}_{bc} = \tfrac12 (c^a{}_{bc} + c_b{}^a{}_c + c_c{}^a{}_b ) - \delta^a{}_b\alpha_c - \delta^a{}_c\alpha_b + \alpha^a\delta_{bc}, \label{eq:gamma abc}
\end{equation}
are those of the Weyl--Levi--Civita connection on the cuts. Within the same set of hypotheses, the diffeomorphism parameters then obey
\begin{equation}
	\bfupsilon(f) = \tfrac12 \mathcal D_a Y^a,\quad \bfupsilon(Y^a) = 0,\quad 2\mathcal D_{(a}Y_{b)} = \mathcal D_c Y^c\delta_{ab}, \label{eq:def bms}
\end{equation}
which define an element of $\mathfrak{bms}_4$, the BMS algebra in four dimensions \cite{Bondi:1962px,Sachs:1962zza,Sachs:1962wk}. The gauge-fixed variation \eqref{eq:gauge fixed var} then induces a representation of this algebra on the boundary and acts therefore as expected from the bulk analysis \cite{Barnich:2010eb,Compere:2018ylh} on dynamical variables as, \textit{e.g.}, the asymptotic shear \footnote{Eq. \eqref{eq:delta Cab} reproduces exactly Eq. (4.56) of \cite{Barnich:2010eb}, up to three amendements: the sign convention in the variation, the fact that $l = 0$, as it corresponds to $\theta$ here.} 
\begin{equation}
	\delta_\bfxi \mathcal C_{ab} = f\bfupsilon(\mathcal C_{ab}) + (\mathscr L_Y \mathcal C)_{ab} - \bfupsilon(f)\mathcal C_{ab} - 2\mathcal D_{\langle a}\mathcal D_{b\rangle} f. \label{eq:delta Cab}
\end{equation}
Importantly, the inhomogeneous piece above betrays that the transverse tensor $\mathcal C_{ab}$ appears in the spacelike connection under Carroll boosts.

The Carroll and Weyl-covariant flux-balance laws \eqref{eq:flux balance general} now read explicitly in the BMS frame as
\begin{subequations}
\begin{align}
	\bfupsilon(\Pi) &= \mathcal D_a\mathcal D_b \Omega_0{}^{ba} + \bfupsilon\big(\mathcal D_a \Omega_0{}^{0a}\big) - \tfrac12 \mathcal N_{ab} \Omega_0{}^{ab}, \label{eq:fluxbal Pi} \\
	\bfupsilon(P_a) &= \tfrac12 \mathcal D_a \Pi - \mathcal D^b \Pi_{\langle ab\rangle} + \tfrac12 \bfupsilon\big(\mathcal D_a \Omega_0{}^{00} + \mathcal D_a \Omega_b{}^{0b}\big) \nonumber \\
	&\quad + \tfrac12 \mathcal D_a\mathcal D_b\big( \Omega_0{}^{b0} + \Omega_c{}^{bc} \big) + \bfupsilon\big(\mathcal D^b\Omega_{[a|0|b]}\big) + \mathcal R \Omega_{[a}{}^b{}_{b]} \nonumber \\ 
	&\quad + \mathcal D^b\mathcal D^c \Omega_{[a|c|b]} + \tfrac12 \mathcal N_{ab}\Omega_0{}^{0b} - \tfrac12 \mathcal D_{[a}\mathcal C_{b]c} \,\Omega_0{}^{bc}, \label{eq:fluxbal P a}
\end{align}
\end{subequations}
where the constraints \eqref{eq:energy flux} and \eqref{eq:constraints rot weyl} have been extensively used and $\Omega_A{}^{CB} \equiv \htheta^C(\bm\Omega_A{}^B)$ denote the components of the hypermomentum. Note that the news tensor has been defined as $\mathcal N_{ab}\equiv -2 \mathcal  R^0{}_{a0b}$ \cite{Campoleoni:2023fug}. This definition coincides with the retarded-time-derivative of the shear, up to geometrical contributions ensuring that its Weyl-covariance. A straightforward computation shows that upon identifying the hypermomenta as
\begin{equation}
\begin{split}
	\bm\Omega_0{}^0 &= \bm 0,\quad \bm\Omega_{ab} = \bm\Omega_{[ab]} = \mathcal D_{[a} \mathcal C^c{}_{b]}\hframe_c, \\
	 \bm\Omega_0{}^a &= \big(\mathcal N^{ab} + \tfrac12\mathcal R \delta^{ab}\big)\hframe_b,
\end{split}
	\label{eq:hypermomenta}
\end{equation}
the equations \eqref{eq:fluxbal Pi} and \eqref{eq:fluxbal P a} reproduce the BMS flux-balance equations \eqref{eq:bondi mass loss} and \eqref{eq:bondi am loss}, provided that the Carrollian momenta are given in terms of the bulk gravitational data as
\begin{subequations}\label{eq:holo dico}
\begin{align}
	\Pi &= 4M, \quad P_a = 2N_a + \tfrac{1}{16}\mathcal D_a\big(\mathcal C^{bc} \mathcal C_{bc}\big), \label{eq:holo dico 1} \\ 
	\Pi^a &= -\mathcal D_b\mathcal N^{ab} - \tfrac12 \mathcal D^a\mathcal R, \label{eq:holo dico 2} \\
	\Pi_{ab} &= \mathcal D^c\mathcal D_{[a} \mathcal C_{b]c} - \tfrac12 \mathcal N_{[a}{}^c\mathcal C_{b]c} -\tfrac14 \mathcal R\mathcal C_{ab} - 2 M \delta_{ab} . \label{eq:holo dico 3}
\end{align}
\end{subequations}
Eqs. \eqref{eq:hypermomenta} and \eqref{eq:holo dico} constitute our proposal for the flat-space holographic dictionary for gravitational fields. In particular, the boundary momenta are identified, up to radiative decorations, with the bulk gravitational momenta $M, N_a$, which appear at third subleading order in the conformally rescaled metric, exactly as in the AdS case \cite{Henningson:1998gx,Balasubramanian:1999re}. This completes our boundary-intrinsic and first-principles derivation of the BMS evolution equations for asymptotically flat gravitational fields, which is an important step forward in understanding holographic duality in this context.

\section{Discussion}

To conclude our analysis, some remarks are in order. First of all, the holographic dictionary \eqref{eq:holo dico} allows to recover the BMS charges computed via covariant-phase-space methods from a purely boundary analysis. Indeed, taking the identifications \eqref{eq:holo dico 1} into account, the boundary Noether charges on any cut $\Sigma$ of $\mathscr I$ are given by
\begin{align}
	\mathcal Q_{(f,Y)} &= \frac{1}{16\pi G}\int_{\Sigma}\bfmu_{\Sigma}\left(f \Pi + Y^a P_a \right) \label{eq:bms charges} \\
	&= \frac{1}{16\pi G} \int_{\Sigma}\bfmu_{\Sigma}\left(4 f M + 2 Y^a N_a + \tfrac{1}{16}Y^a \mathcal D_a(\mathcal C_{bc}\mathcal C^{bc}) \right),  \nonumber
\end{align}
where $\bfmu_\Sigma\equiv \htheta^1\wedge\htheta^2$ is the volume form on $\Sigma$, which corresponds exactly to the Barnich--Troessaert charges \cite{Barnich:2011mi}. Whether these Carrollian momenta correspond to the Brown--York energy--momentum tensor of null infinity \cite{Chandrasekaran:2018aop}, as it is the case in AdS \cite{Balasubramanian:1999re,Skenderis:2000in}, is a relevant question, which deserves further investigation. Furthermore, the rationale behind the fact that our analysis lands on the ``bare'' BMS momenta of \cite{Barnich:2011mi} and not their covariant avatars, conveniently encoded by the Newman--Penrose coefficients $\Psi^0_1$ and $\Psi^0_2$ and promoted into preferred BMS charge prescription in \cite{Compere:2020lrt,Freidel:2021fxf,Freidel:2021cjp,Campoleoni:2023fug,Kmec:2024nmu},\footnote{See also \cite{Rignon-Bret:2024gcx,Rignon-Bret:2024mef} for another proposal for the covariant BMS charges derived from a Wald--Zoupas prescription \cite{Wald:1999wa} and based on Geroch's supermomentum \cite{Geroch:1977big}.} is that we have also considered the ``bare'' momenta coming from the variation \eqref{eq:delta S} without any improvement. The right-hand sides in Eqs. \eqref{eq:flux balance general} contain some terms that can be recast as divergences, which are able to nourish the ``bare'' momenta with radiative contributions and transform them into their covariant counterpart. 

As a conclusion, our work sheds a new light on the properties of the putative dual field theory in flat-space holography. Firstly, it shows that the asymptotic Einstein equations can be derived in a holographic manner with a very minimalistic holographic dictionary at disposal. In particular, the use of hypermomenta techniques in this setting offers a geometric perspective on the interplay between the dual theory and the radiative sources, further supporting the heuristic approach of \cite{Donnay:2022wvx}. Secondly, it allows for a genuinely Carrollian understanding of asymptotically flat gravity and its presumed holographic realisation without relying on limits from AdS \cite{Campoleoni:2023fug}. Finally, it paves the way towards grasping the microscopic structure of the dual theory, namely the sub-sector governing the sources: at this stage, it appears clear that the latter should encompass a theory for the connection itself, and whose variation with respect to boundary geometric data is constrained by the form \eqref{eq:hypermomenta} of the hypermomenta. Unravelling this theory could be of relevance for Carrollian amplitudes.

\section*{Acknowledgements}
We would like to thank Ismaël Ahlouche, Mathieu Beauvillain, Andrea Campoleoni, Marc Geiller, Daniel Grumiller, Damianos Iosifidis, Lionel Mason, Anastasios C. Petkou, Antoine Rignon--Bret, Romain Ruzziconi, Konstantinos Siampos and Céline Zwikel for useful discussions. Adrien Fiorucci acknowledges support from the FWF (Austria) during the early phases of this work (projects P~33789, P~36619). The work of Simon Pekar was supported by the \emph{Fonds Friedmann} run by the \emph{Fondation de l'\'Ecole polytechnique} until 31 October 2024. Since 1 November 2024, Simon is supported by the \textit{European Research Council (ERC) Project 101076737 -- CeleBH}. Views and opinions expressed are however those of the author only and do not necessarily reflect those of the European Union or the European Research Council. Neither the European Union nor the granting authority can be held responsible for them. Simon is also partially supported by INFN Iniziativa Specifica ST\& FI. The work of Matthieu Vilatte is supported by the F.R.S.-FNRS (Belgium) under the Grant No. T.0047.24. Matthieu is grateful to the \textit{Centre de Physique Théorique de l'\'Ecole polytechnique}, and in particular Jean-René Chazottes, for financial support during summer 2024. Finally, we would like to acknowledge support from the \textit{Partenariat Hubert Curien France/Grèce}, project entitled \textit{Symétries
asymptotiques en gravitation, théories des champs conformes et holographie}. We are also grateful to the \textit{Laboratory of Theoretical Physics} of the \textit{Aristotle University of Thessaloniki} and the \textit{Mathematical Institute} of the \textit{University of Oxford} for the organisation of brainstorming meetings,  where parts of this work have been carried out and its fundamental ideas have been discussed. 

\appendix

\section{Transformation laws}

We collect here the transformation laws of the structure functions and the connection one-forms under infinitesimal local transformations and diffeomorphisms on the boundary. We keep denoting by $\lambda_a$ and $r^a{}_b$ the infinitesimal parameters of local Carroll rotations and boosts, which act respectively as \eqref{eq:rotation} and \eqref{eq:boost} on the Carroll--Cartan basis. 

As our definition of the structure functions is still given by Eq. \eqref{eq:non holonomy}, we can check that
\begin{subequations}
	\begin{align}
		\delta_\lambda \varphi_a &= \lambda_c c^c{}_a + \bfupsilon(\lambda_a), \\
		\delta_\lambda \varpi_{ab} &= \hframe_{[a}(\lambda_{b]}) + \lambda_{[a}\varphi_{b]} - \tfrac12 \lambda_c c^c{}_{ab}, \\
		\delta_\lambda c^a{}_b &= 0, \\ \delta_\lambda c^c{}_{ab} &= 2 c^c{}_{[a}\lambda_{b]} ,
	\end{align}
\end{subequations}
under local Carroll boosts,
\begin{subequations}
	\begin{align}
		\delta_r \varphi_a &= r_a{}^b\varphi_b, \\
		 \delta_r \varpi_{ab} &= r_a{}^c\varpi_{cb} + r_b{}^c\varpi_{ac}, \\
		 \delta_r c^a{}_b &= r^a{}_c c^c{}_b + r_b{}^c c^a{}_c + \bfupsilon(r^a{}_b), \label{eq:delta r cab} \\
		\delta_r c^a{}_{bc} &= r^a{}_d c^d{}_{bc} + r_b{}^d c^a{}_{dc} + r_c{}^d c^a{}_{bd} - 2 \hframe_{[b}(r^a{}_{c]}),
	\end{align}
\end{subequations}
under local rotations and finally
\begin{subequations}
	\begin{align}
		\delta_B \varphi_a &= B\varphi_a - \hframe_a(B), \\ 
		\delta_B c^a{}_b &= B c^a{}_b - \delta^a{}_b\bfupsilon(B), \\
		\delta_B \varpi_{ab} &= B\varpi_{ab}, \\
		\delta_B c^a{}_{bc} &= B c^a{}_{bc} - 2 \delta^a{}_{[b}\hframe_{c]}(B) \label{eq:delta B cab}
	\end{align}
\end{subequations}
under Weyl rescalings. From Eq. \eqref{eq:delta B cab}, it can be observed that the gauge choice $\theta = c^a{}_a = 0$ implies the reduction of Weyl transformations to time-invariant parameters, $\bfupsilon(B) = 0$. The additional choice $c_{[ab]} = 0$ reduces the local symmetry group to time-invariant rotations, $\bfupsilon(r_{ab})=0$, as indicated by Eq. \eqref{eq:delta r cab}. Assuming this, the transverse tensor $c_{ab}$ identically vanishes on account of asymptotic Einstein equations, as explained in the main text.

Let us now consider a general connection $\nabla$ on $T\mathscr I$. Looking at the transformation laws of the related connection one-forms $\bfomega^A{}_B$, we find
\begin{subequations}\label{eq:omega boost}
\begin{align}
	\delta_\lambda \bfomega^0{}_0 &= -\lambda_a \bfomega^a{}_0, & \delta_\lambda \bfomega^a{}_0 &= \bm 0, \\
	\delta_\lambda \bfomega^0{}_a &= {\bm\nabla} \lambda_a + \lambda_a \bfomega^0{}_0, & \delta_\lambda \bfomega^a{}_b &= \lambda_b \bfomega^a{}_0,  \label{eq:omega 0a boost}
\end{align}
\end{subequations}
under local Carroll boosts,
\begin{subequations}
\begin{align}
	\delta_r \bfomega^0{}_0 &= \bm 0, & \delta_r  \bfomega^a{}_0 &= r^a{}_b \bfomega^b{}_0, \\ 
	\delta_r \bfomega^0{}_a &= r_a{}^b \bfomega^0{}_b, & \delta_r \bfomega^a{}_b &= \bm\nabla r_b{}^a, \label{eq:omega ab rot}
\end{align}
\end{subequations}
under local rotations, and
\begin{subequations}
\begin{align}
	\delta_B \bfomega^0{}_0 &= \D B, & \delta_B  \bfomega^a{}_0 &= \bm 0 \\ 
	\delta_B \bfomega^0{}_a &= \bm 0, & \delta_B \bfomega^a{}_b &= \delta^a{}_b \D B,
\end{align}
\end{subequations}
under Weyl rescalings. In Eqs. \eqref{eq:omega 0a boost} and \eqref{eq:omega ab rot}, the bold symbol $\bm\nabla(\cdot) = \bftau\nabla_{\bfupsilon}(\cdot)+\htheta^a\nabla_a(\cdot)$ represents the covariant exterior derivative. The connection one-forms $\bfomega^0{}_0$ (together with $\bfomega^a{}_a$), $\bfomega^0{}_a$ and $\bfomega_{[ab]}$ are respectively recognised as Weyl, boost and rotation connections. Furthermore, $\bfomega^a{}_0$ transforms homogeneously under all the local transformations. The latter can always be set to zero without any restriction, as it can be shown to form the connection for Galilean boosts,
\begin{equation}
	\delta_w\bfupsilon = w^a \hframe_a,\quad \delta_w \hframe_a = \bm 0,\quad \delta_w\bftau = \bm 0,\quad \delta_w\htheta^a = - w^a \bftau, \label{eq:galilean}
\end{equation}
that mirror the operation of \eqref{eq:boost} and are absent in Carroll relativity. In the same vein, we observe that $\bfomega_{\langle ab \rangle}$ also transforms homogeneously under all the local Carroll and Weyl transformations and can always be safely set to zero.

\section{Derivation of Carroll flux-balance laws}
\label{app:deriv}
In this Appendix, we provide some details about the derivation of Eqs. \eqref{eq:flux balance general}. We start with the general variation \eqref{eq:delta S} of the boundary action, we evaluate it on a boundary diffeomorphism $\bfxi$, and then demand $S$ to be invariant under its action when the fields $\Phi$ are on-shell,
\begin{equation}
	\mathscr L_\bfxi S \approx \frac{1}{16\pi G}\int_\mathscr{I} \bfmu \big(\mathscr L_\bfxi \htheta^A (\text{\bf T}_A) + \mathscr L_\bfxi \bfomega^A{}_B (\bm\Omega_A{}^B) \big) \approx 0. \label{eq:delta S diffeo}
\end{equation}
The Lie derivatives that appear in the integrand are computed using Cartan's magic formula as
\begin{align}
	\mathscr L_\bfxi \htheta^A &= \big(\hframe_B(\xi^A) + C^A{}_{BC}\xi^C \big)\htheta^B, \\
	\mathscr L_\bfxi \bfomega^A{}_B &= \big(\xi^C\hframe_C(\omega^A{}_{DB}) + \omega^A{}_{CB}\hframe_D(\xi^C) + \omega^A{}_{CB}C^C{}_{DE}\xi^E \big)\htheta^D \nonumber
\end{align}
where $[\hframe_A,\hframe_B] = C^C{}_{AB}\hframe_C$ and $\bfomega^A{}_B = \omega^A{}_{CB}\htheta^C$. At the moment, $\bfomega^A{}_B$ are the one-forms of a general connection on the boundary. To isolate the diffeomorphism parameter, we need an inverse Leibniz rule on the differential operator $\hframe_A$, which is again derived using Cartan's magic formula and reads
\begin{equation}
	\bfmu f \hframe_A(g) = \D(\dots) - \bfmu \big(g \hframe_A(f) - f g C^B{}_{AB}\big)
	\label{eq:ipp}
\end{equation}
for all functions $f,g$ on $\mathscr I$. Since the boundary metric admits no inverse, the computation should be done separately for $A = 0$ and $A = a$. The final result can be recast covariantly with respect to the Carroll--Cartan frame as in Eq. \eqref{eq:ipp}. 

Next, we shall keep in mind that, after taking the variation, the result needs to be evaluated for the boundary connection $\nabla$ which is torsion-free and obeys \eqref{eq:strong weyl}. These conditions are solved by
\begin{equation}
	\bfomega^0{}_0 = -\bm\upalpha,\quad \bfomega^a{}_0 = \bm 0, \quad \bfomega_{(ab)} = -\bm\upalpha \delta_{ab}, \label{eq:robocop}
\end{equation}
and the other connection coefficients are determined using the fact that the torsion vanishes, which, by virtue of first Cartan structure equation, implies
\begin{equation}
	\D\htheta^A + \bfomega^A{}_B\wedge \htheta^B = \bm 0 \quad\Rightarrow\quad \omega^A{}_{[BC]} = \tfrac12 C^A{}_{BC}.
\end{equation}
We then find
\begin{subequations}
\begin{align}
	\omega^0{}_{0a} &= \varphi_a-\alpha_a,& \omega^0{}_{ab} &= \beta_{(ab)} + \varpi_{ab}, \\ 
	\omega^a{}_{0b} &= -c^a{}_b, & 
	\omega^a{}_{bc} &= \gamma^a{}_{bc} = \eqref{eq:gamma abc},
\end{align}
\end{subequations}
where $\beta_{(ab)}$ are utterly free data, as we explained in the main text. Note that the identities $\omega^B{}_{AB} = -3\alpha_A$ and $\omega_{(a|C|b)} = -\alpha_C \delta_{ab}$ hold, which is verifiable by direct computation. The condition \eqref{eq:delta S diffeo} can then be worked out to get
\begin{equation}
\begin{split}
	\int_{\mathscr I}\bfmu \,\xi^A\, &\Big(\pmb{\mathcal D}\cdot \text{\bf T}_A - \mathcal R^C{}_{BAD} \Omega_C{}^{DB} + \omega^C{}_{AB}\Delta^B{}_C\Big) \approx 0,
\end{split} \label{eq:delta xi developed}
\end{equation}
where the boundary term has been discarded and $\xi^A$ is a general function. The last term involves the combination \mbox{$\Delta^B{}_C\equiv T^B{}_C + \pmb{\mathcal D}\cdot \bm\Omega_C{}^B$} and can thus be shown to vanish if one requires local Carroll and Weyl invariance. Indeed,
\begin{equation}
	\begin{split}
		\omega^C{}_{AB}\Delta^B{}_C &= -\alpha_A \Delta^0{}_0 + \omega^0{}_{Ab}\Delta^b{}_0 + \omega^c{}_{Ab}\Delta^b{}_c \\
		&= \omega_{cAb}\Delta^{[bc]} - \alpha_A\big(\Delta^0{}_0 + \Delta^a{}_a\big) = 0.
	\end{split}
\end{equation}
The first equality uses the second condition in Eq. \eqref{eq:robocop}, the second one uses the condition \eqref{eq:energy flux}, $\Delta^a{}_0 = 0$, coming from Carroll-boost invariance, separates the spacelike indices $(b,c)$ into symmetric and antisymmetric parts and uses the third condition in Eq. \eqref{eq:robocop}. The result vanishes by virtue of the constraints \eqref{eq:constraints rot weyl} imposed by local rotation and Weyl invariance, \textit{i.e.}, $\Delta^A{}_A = 0$ and $\Delta_{[ab]}=0$. Therefore, Eq. \eqref{eq:delta xi developed} implies the general flux-balance laws displayed in Eq. \eqref{eq:flux balance general}.

%\bibliographystyle{style}
%\bibliography{carrollrefs}

\begin{thebibliography}{100}

\bibitem{Leblond}
J.-M. L\'evy-Leblond, \emph{{Une nouvelle limite non-relativiste du groupe de
  Poincar\'e}}, A. Inst. Henri Poincar\'e III 1
(1965)
% .

\bibitem{SenGupta:1966qer}
N.~D. Sen~Gupta, \emph{{On an analogue of the Galilei group}}, Nuovo Cim. A
  {\bf 44} (1966), no.~2,
512--517
% .

\bibitem{Duval:2014uva}
C.~Duval, G.~W. Gibbons  and P.~A. Horvathy, \emph{{Conformal Carroll groups
  and BMS symmetry}}, Class. Quant. Grav. {\bf 31} (2014) 092001,
\href{http://www.arXiv.org/abs/1402.5894}{{1402.5894}}
% .

\bibitem{Duval:2014lpa}
C.~Duval, G.~W. Gibbons  and P.~A. Horvathy, \emph{{Conformal Carroll groups}},
  J. Phys. A {\bf 47} (2014), no.~33, 335204,
\href{http://www.arXiv.org/abs/1403.4213}{{1403.4213}}
% .

\bibitem{Ciambelli:2019lap}
L.~Ciambelli, R.~G. Leigh, C.~Marteau  and P.~M. Petropoulos, \emph{{Carroll
  Structures, Null Geometry and Conformal Isometries}}, Phys. Rev. D {\bf 100}
  (2019), no.~4, 046010,
\href{http://www.arXiv.org/abs/1905.02221}{{1905.02221}}
% .

\bibitem{Campoleoni:2021blr}
A.~Campoleoni and S.~Pekar, \emph{{Carrollian and Galilean conformal
  higher-spin algebras in any dimensions}}, JHEP {\bf 02} (2022) 150,
\href{http://www.arXiv.org/abs/2110.07794}{{2110.07794}}
% .

\bibitem{Chen:2021xkw}
B.~Chen, R.~Liu  and Y.-f. Zheng, \emph{{On higher-dimensional Carrollian and
  Galilean conformal field theories}}, SciPost Phys. {\bf 14} (2023), no.~5,
  088,
\href{http://www.arXiv.org/abs/2112.10514}{{2112.10514}}
% .

\bibitem{Donnay:2022aba}
L.~Donnay, A.~Fiorucci, Y.~Herfray  and R.~Ruzziconi, \emph{{Carrollian
  Perspective on Celestial Holography}}, Phys. Rev. Lett. {\bf 129} (2022),
  no.~7, 071602,
\href{http://www.arXiv.org/abs/2202.04702}{{2202.04702}}
% .

\bibitem{Donnay:2022wvx}
L.~Donnay, A.~Fiorucci, Y.~Herfray  and R.~Ruzziconi, \emph{{Bridging
  Carrollian and celestial holography}}, Phys. Rev. D {\bf 107} (2023), no.~12,
  126027,
\href{http://www.arXiv.org/abs/2212.12553}{{2212.12553}}
% .

\bibitem{Rivera-Betancour:2022lkc}
D.~Rivera-Betancour and M.~Vilatte, \emph{{Revisiting the Carrollian scalar
  field}}, Phys. Rev. D {\bf 106} (2022), no.~8, 085004,
\href{http://www.arXiv.org/abs/2207.01647}{{2207.01647}}
% .

\bibitem{Baiguera:2022lsw}
S.~Baiguera, G.~Oling, W.~Sybesma  and B.~T. S\o{}gaard, \emph{{Conformal
  Carroll Scalars with Boosts}},
\href{http://www.arXiv.org/abs/2207.03468}{{2207.03468}}
% .

\bibitem{Bekaert:2022oeh}
X.~Bekaert, A.~Campoleoni  and S.~Pekar, \emph{{Carrollian conformal scalar as
  flat-space singleton}}, Phys. Lett. B {\bf 838} (2023) 137734,
\href{http://www.arXiv.org/abs/2211.16498}{{2211.16498}}
% .

\bibitem{Nguyen:2023vfz}
K.~Nguyen and P.~West, \emph{{Carrollian Conformal Fields and Flat
  Holography}}, Universe {\bf 9} (2023), no.~9, 385,
\href{http://www.arXiv.org/abs/2305.02884}{{2305.02884}}
% .

\bibitem{Bekaert:2024itn}
X.~Bekaert, A.~Campoleoni  and S.~Pekar, \emph{{Holographic Carrollian
  conformal scalars}}, JHEP {\bf 05} (2024) 242,
\href{http://www.arXiv.org/abs/2404.02533}{{2404.02533}}
% .

\bibitem{Chen:2024voz}
B.~Chen, H.~Sun  and Y.-f. Zheng, \emph{{Quantization of Carrollian conformal
  scalar theories}}, Phys. Rev. D {\bf 110} (2024), no.~12, 125010,
\href{http://www.arXiv.org/abs/2406.17451}{{2406.17451}}
% .

\bibitem{Trautman:1958zdi}
A.~Trautman, \emph{{Radiation and Boundary Conditions in the Theory of
  Gravitation}}, Bull. Acad. Pol. Sci. Ser. Sci. Math. Astron. Phys. {\bf 6}
  (1958), no.~6, 407--412,
\href{http://www.arXiv.org/abs/1604.03145}{{1604.03145}}
% .

\bibitem{Bondi:1962px}
H.~Bondi, M.~G.~J. van~der Burg  and A.~W.~K. Metzner, \emph{{Gravitational
  waves in general relativity. 7. Waves from axisymmetric isolated systems}},
  Proc. Roy. Soc. Lond. {\bf A269} (1962)
21
%%CITATION = PRSLA,A269,21;%%.

\bibitem{Sachs:1962wk}
R.~K. Sachs, \emph{{Gravitational waves in general relativity. 8. Waves in
  asymptotically flat space-times}}, Proc. Roy. Soc. Lond. {\bf A270} (1962)
103--126
%%CITATION = PRSLA,A270,103;%%.

\bibitem{Sachs:1962zza}
R.~Sachs, \emph{{Asymptotic symmetries in gravitational theory}}, Phys. Rev.
  {\bf 128} (1962)
2851--2864
%%CITATION = PHRVA,128,2851;%%.

\bibitem{Tamburino:1966zz}
L.~A. Tamburino and J.~H. Winicour, \emph{{Gravitational Fields in Finite and
  Conformal Bondi Frames}}, Phys. Rev. {\bf 150} (1966)
1039--1053
% .

\bibitem{Barnich:2010eb}
G.~Barnich and C.~Troessaert, \emph{{Aspects of the BMS/CFT correspondence}},
  JHEP {\bf 05} (2010) 062,
\href{http://www.arXiv.org/abs/1001.1541}{{1001.1541}}
%%CITATION = ARXIV:1001.1541;%%.

\bibitem{Ruzziconi:2020cjt}
R.~Ruzziconi, {\em {On the Various Extensions of the BMS Group}}.
\newblock PhD thesis, U. Brussels, 2020.
\newblock
\href{http://www.arXiv.org/abs/2009.01926}{{2009.01926}}.
\newblock
% .

\bibitem{Fiorucci:2021pha}
A.~Fiorucci, {\em {Leaky covariant phase spaces: Theory and application to
  \ensuremath{\Lambda}-BMS symmetry}}.
\newblock PhD thesis, Brussels U., Intl. Solvay Inst., Brussels, 2021.
\newblock
\href{http://www.arXiv.org/abs/2112.07666}{{2112.07666}}.
\newblock
% .

\bibitem{Freidel:2021fxf}
L.~Freidel, R.~Oliveri, D.~Pranzetti  and S.~Speziale, \emph{{The Weyl BMS
  group and Einstein\textquoteright{}s equations}}, JHEP {\bf 07} (2021) 170,
\href{http://www.arXiv.org/abs/2104.05793}{{2104.05793}}
% .

\bibitem{Geiller:2022vto}
M.~Geiller and C.~Zwikel, \emph{{The partial Bondi gauge: Further enlarging the
  asymptotic structure of gravity}}, SciPost Phys. {\bf 13} (2022) 108,
\href{http://www.arXiv.org/abs/2205.11401}{{2205.11401}}
% .

\bibitem{Geiller:2024ryw}
M.~Geiller, A.~Laddha  and C.~Zwikel, \emph{{Symmetries of the gravitational
  scattering in the absence of peeling}}, JHEP {\bf 12} (2024) 081,
\href{http://www.arXiv.org/abs/2407.07978}{{2407.07978}}
% .

\bibitem{Campoleoni:2023fug}
A.~Campoleoni, A.~Delfante, S.~Pekar, P.~M. Petropoulos, D.~Rivera-Betancour
  and M.~Vilatte, \emph{{Flat from anti de Sitter}}, JHEP {\bf 12} (2023) 078,
\href{http://www.arXiv.org/abs/2309.15182}{{2309.15182}}
% .

\bibitem{Geroch:1977big}
R.~Geroch, \emph{{Asymptotic Structure of Space-Time}}, in {\em {Symposium on
  Asymptotic Structure of Space-Time}}.
\newblock
1977.
\newblock
% .

\bibitem{Ashtekar:1981bq}
A.~Ashtekar and M.~Streubel, \emph{{Symplectic Geometry of Radiative Modes and
  Conserved Quantities at Null Infinity}}, Proc. Roy. Soc. Lond. {\bf A376}
  (1981)
585--607
%%CITATION = PRSLA,A376,585;%%.

\bibitem{Wald:1999wa}
R.~M. Wald and A.~Zoupas, \emph{{A General definition of `conserved quantities'
  in general relativity and other theories of gravity}}, Phys. Rev. {\bf D61}
  (2000) 084027,
\href{http://www.arXiv.org/abs/gr-qc/9911095}{{gr-qc/9911095}}
%%CITATION = GR-QC/9911095;%%.

\bibitem{Barnich:2011mi}
G.~Barnich and C.~Troessaert, \emph{{BMS charge algebra}}, JHEP {\bf 12} (2011)
  105,
\href{http://www.arXiv.org/abs/1106.0213}{{1106.0213}}
% .

\bibitem{Troessaert:2015nia}
C.~Troessaert, \emph{{Hamiltonian surface charges using external sources}}, J.
  Math. Phys. {\bf 57} (2016), no.~5, 053507,
\href{http://www.arXiv.org/abs/1509.09094}{{1509.09094}}
% .

\bibitem{Wieland:2020gno}
W.~Wieland, \emph{{Null infinity as an open Hamiltonian system}}, JHEP {\bf 04}
  (2021) 095,
\href{http://www.arXiv.org/abs/2012.01889}{{2012.01889}}
% .

\bibitem{Fiorucci:talk}
A.~Fiorucci, \emph{Hamiltonian Mechanics with External Sources}, Talk at 2nd
  Carroll Workshop, Mons, September 2023
(based on unpublished results with G. Barnich and R. Ruzziconi)
% .

\bibitem{Mason:2023mti}
L.~Mason, R.~Ruzziconi  and A.~Yelleshpur~Srikant, \emph{{Carrollian amplitudes
  and celestial symmetries}}, JHEP {\bf 05} (2024) 012,
\href{http://www.arXiv.org/abs/2312.10138}{{2312.10138}}
% .

\bibitem{Liu:2024nfc}
W.-B. Liu, J.~Long  and X.-Q. Ye, \emph{{Feynman rules and loop structure of
  Carrollian amplitudes}}, JHEP {\bf 05} (2024) 213,
\href{http://www.arXiv.org/abs/2402.04120}{{2402.04120}}
% .

\bibitem{Stieberger:2024shv}
S.~Stieberger, T.~R. Taylor  and B.~Zhu, \emph{{Carrollian Amplitudes from
  Strings}}, JHEP {\bf 04} (2024) 127,
\href{http://www.arXiv.org/abs/2402.14062}{{2402.14062}}
% .

\bibitem{Ruzziconi:2024zkr}
R.~Ruzziconi, S.~Stieberger, T.~R. Taylor  and B.~Zhu, \emph{{Differential
  equations for Carrollian amplitudes}}, JHEP {\bf 09} (2024) 149,
\href{http://www.arXiv.org/abs/2407.04789}{{2407.04789}}
% .

\bibitem{Kraus:2024gso}
P.~Kraus and R.~M. Myers, \emph{{Carrollian partition functions and the flat
  limit of AdS}}, JHEP {\bf 01} (2025) 183,
\href{http://www.arXiv.org/abs/2407.13668}{{2407.13668}}
% .

\bibitem{Liu:2024llk}
W.-B. Liu, J.~Long, H.-Y. Xiao  and J.-L. Yang, \emph{{On the definition of
  Carrollian amplitudes in general dimensions}}, JHEP {\bf 11} (2024) 027,
\href{http://www.arXiv.org/abs/2407.20816}{{2407.20816}}
% .

\bibitem{Ruzziconi:2024kzo}
R.~Ruzziconi and A.~Saha, \emph{{Holographic Carrollian currents for massless
  scattering}}, JHEP {\bf 01} (2025) 169,
\href{http://www.arXiv.org/abs/2411.04902}{{2411.04902}}
% .

\bibitem{Kraus:2025wgi}
P.~Kraus and R.~M. Myers, \emph{{Carrollian Partition Function for Bulk
  Yang-Mills Theory}},
\href{http://www.arXiv.org/abs/2503.00916}{{2503.00916}}
% .

\bibitem{Nguyen:2025sqk}
K.~Nguyen and J.~Salzer, \emph{{Operator Product Expansion in Carrollian CFT}},
\href{http://www.arXiv.org/abs/2503.15607}{{2503.15607}}
% .

\bibitem{Alday:2024yyj}
L.~F. Alday, M.~Nocchi, R.~Ruzziconi  and A.~Yelleshpur~Srikant,
  \emph{{Carrollian Amplitudes from Holographic Correlators}},
\href{http://www.arXiv.org/abs/2406.19343}{{2406.19343}}
% .

\bibitem{Barnich:2021dta}
G.~Barnich and R.~Ruzziconi, \emph{{Coadjoint representation of the BMS group
  on celestial Riemann surfaces}},
\href{http://www.arXiv.org/abs/2103.11253}{{2103.11253}}
% .

\bibitem{Barnich:2022bni}
G.~Barnich, K.~Nguyen  and R.~Ruzziconi, \emph{{Geometric action for extended
  Bondi-Metzner-Sachs group in four dimensions}}, JHEP {\bf 12} (2022) 154,
\href{http://www.arXiv.org/abs/2211.07592}{{2211.07592}}
% .

\bibitem{Ciambelli:2018xat}
L.~Ciambelli, C.~Marteau, A.~C. Petkou, P.~M. Petropoulos  and K.~Siampos,
  \emph{{Covariant Galilean versus Carrollian hydrodynamics from relativistic
  fluids}}, Class. Quant. Grav. {\bf 35} (2018), no.~16, 165001,
\href{http://www.arXiv.org/abs/1802.05286}{{1802.05286}}
% .

\bibitem{Ciambelli:2018wre}
L.~Ciambelli, C.~Marteau, A.~C. Petkou, P.~M. Petropoulos  and K.~Siampos,
  \emph{{Flat holography and Carrollian fluids}}, JHEP {\bf 07} (2018) 165,
\href{http://www.arXiv.org/abs/1802.06809}{{1802.06809}}
% .

\bibitem{Campoleoni:2018ltl}
A.~Campoleoni, L.~Ciambelli, C.~Marteau, P.~M. Petropoulos  and K.~Siampos,
  \emph{{Two-dimensional fluids and their holographic duals}}, Nucl. Phys. B
  {\bf 946} (2019) 114692,
\href{http://www.arXiv.org/abs/1812.04019}{{1812.04019}}
% .

\bibitem{Mittal:2022ywl}
N.~Mittal, P.~M. Petropoulos, D.~Rivera-Betancour  and M.~Vilatte,
  \emph{{Ehlers, Carroll, charges and dual charges}}, JHEP {\bf 07} (2023) 065,
\href{http://www.arXiv.org/abs/2212.14062}{{2212.14062}}
% .

\bibitem{Li:2010dr}
W.~Li and T.~Takayanagi, \emph{{Holography and Entanglement in Flat
  Spacetime}}, Phys. Rev. Lett. {\bf 106} (2011) 141301,
\href{http://www.arXiv.org/abs/1010.3700}{{1010.3700}}
% .

\bibitem{Barnich:2012xq}
G.~Barnich, \emph{{Entropy of three-dimensional asymptotically flat
  cosmological solutions}}, JHEP {\bf 10} (2012) 095,
\href{http://www.arXiv.org/abs/1208.4371}{{1208.4371}}
% .

\bibitem{Barnich:2012rz}
G.~Barnich, A.~Gomberoff  and H.~A. Gonz\'alez, \emph{{Three-dimensional
  Bondi-Metzner-Sachs invariant two-dimensional field theories as the flat
  limit of Liouville theory}}, Phys. Rev. D {\bf 87} (2013), no.~12, 124032,
\href{http://www.arXiv.org/abs/1210.0731}{{1210.0731}}
% .

\bibitem{Barnich:2013yka}
G.~Barnich and H.~A. Gonzalez, \emph{{Dual dynamics of three dimensional
  asymptotically flat Einstein gravity at null infinity}}, JHEP {\bf 05} (2013)
  016,
\href{http://www.arXiv.org/abs/1303.1075}{{1303.1075}}
%%CITATION = ARXIV:1303.1075;%%.

\bibitem{Detournay:2014fva}
S.~Detournay, D.~Grumiller, F.~Sch\"oller  and J.~Sim\'on, \emph{{Variational
  principle and one-point functions in three-dimensional flat space Einstein
  gravity}}, Phys. Rev. D {\bf 89} (2014), no.~8, 084061,
\href{http://www.arXiv.org/abs/1402.3687}{{1402.3687}}
% .

\bibitem{Bagchi:2015wna}
A.~Bagchi, D.~Grumiller  and W.~Merbis, \emph{{Stress tensor correlators in
  three-dimensional gravity}}, Phys. Rev. D {\bf 93} (2016), no.~6, 061502,
\href{http://www.arXiv.org/abs/1507.05620}{{1507.05620}}
% .

\bibitem{Jiang:2017ecm}
H.~Jiang, W.~Song  and Q.~Wen, \emph{{Entanglement Entropy in Flat
  Holography}}, JHEP {\bf 07} (2017) 142,
\href{http://www.arXiv.org/abs/1706.07552}{{1706.07552}}
% .

\bibitem{Barnich:2017jgw}
G.~Barnich, H.~A. Gonzalez  and P.~Salgado-Rebolledo, \emph{{Geometric actions
  for three-dimensional gravity}}, Class. Quant. Grav. {\bf 35} (2018), no.~1,
  014003,
\href{http://www.arXiv.org/abs/1707.08887}{{1707.08887}}
% .

\bibitem{Ciambelli:2020eba}
L.~Ciambelli, C.~Marteau, P.~M. Petropoulos  and R.~Ruzziconi, \emph{{Gauges in
  Three-Dimensional Gravity and Holographic Fluids}}, JHEP {\bf 11} (2020) 092,
\href{http://www.arXiv.org/abs/2006.10082}{{2006.10082}}
% .

\bibitem{Campoleoni:2022wmf}
A.~Campoleoni, L.~Ciambelli, A.~Delfante, C.~Marteau, P.~M. Petropoulos  and
  R.~Ruzziconi, \emph{{Holographic Lorentz and Carroll frames}}, JHEP {\bf 12}
  (2022) 007,
\href{http://www.arXiv.org/abs/2208.07575}{{2208.07575}}
% .

\bibitem{Athanasiou:2024lzr}
N.~Athanasiou, P.~M. Petropoulos, S.~M. Schulz  and G.~Taujanskas,
  \emph{{One-dimensional Carrollian fluids. Part I. Carroll-Galilei duality}},
  JHEP {\bf 11} (2024) 012,
\href{http://www.arXiv.org/abs/2407.05962}{{2407.05962}}
% .

\bibitem{Athanasiou:2024ykt}
N.~Athanasiou, P.~M. Petropoulos, S.~Schulz  and G.~Taujanskas,
  \emph{{One-dimensional Carrollian fluids II: $C^1$ blow-up criteria}},
\href{http://www.arXiv.org/abs/2407.05971}{{2407.05971}}
% .

\bibitem{Petropoulos:2024jie}
P.~M. Petropoulos, S.~Schulz  and G.~Taujanskas, \emph{{One-Dimensional
  Carrollian Fluids III: Global Existence and Weak Continuity in $L^\infty$}},
\href{http://www.arXiv.org/abs/2407.05972}{{2407.05972}}
% .

\bibitem{Vogel1965}
W.~O. Vogel, \emph{\"Uber lineare Zusammenh\"ange in singul\"aren Riemannschen
  R\"aumen}, Archiv der Mathematik {\bf 16} (Dec., 1965)
106–116
% .

\bibitem{Datcourt1967}
G.~Daŭtcourt, \emph{Characteristic Hypersurfaces in General Relativity. I},
  Journal of Mathematical Physics {\bf 8} (July, 1967)
1492–1501
% .

\bibitem{jankiewicz1954espaces}
C.~Jankiewicz, \emph{Sur les espaces riemanniens d{\'e}g{\'e}n{\'e}r{\'e}s},
  Bull. Acad. Polon. Sci. Cl. III {\bf 2} (1954)
301
% .

\bibitem{Henneaux:1979vn}
M.~Henneaux, \emph{{Geometry of Zero Signature Space-times}}, Bull. Soc. Math.
  Belg. {\bf 31} (1979)
47--63
% .

\bibitem{Bekaert:2015xua}
X.~Bekaert and K.~Morand, \emph{{Connections and dynamical trajectories in
  generalised Newton-Cartan gravity II. An ambient perspective}}, J. Math.
  Phys. {\bf 59} (2018), no.~7, 072503,
\href{http://www.arXiv.org/abs/1505.03739}{{1505.03739}}
% .

\bibitem{Hartong:2015xda}
J.~Hartong, \emph{{Gauging the Carroll Algebra and Ultra-Relativistic
  Gravity}}, JHEP {\bf 08} (2015) 069,
\href{http://www.arXiv.org/abs/1505.05011}{{1505.05011}}
% .

\bibitem{Ashtekar:1981hw}
A.~Ashtekar, \emph{{Radiative Degrees of Freedom of the Gravitational Field in
  Exact General Relativity}}, J. Math. Phys. {\bf 22} (1981)
2885--2895
% .

\bibitem{Ashtekar:1981sf}
A.~Ashtekar, \emph{{Asymptotic Quantization of the Gravitational Field}}, Phys.
  Rev. Lett. {\bf 46} (1981)
573--576
%%CITATION = PRLTA,46,573;%%.

\bibitem{Hehl:1976kj}
F.~W. Hehl, P.~Von Der~Heyde, G.~D. Kerlick  and J.~M. Nester, \emph{{General
  Relativity with Spin and Torsion: Foundations and Prospects}}, Rev. Mod.
  Phys. {\bf 48} (1976)
393--416
% .

\bibitem{Iosifidis:2019dua}
D.~Iosifidis, {\em {Metric-Affine Gravity and Cosmology/Aspects of Torsion and
  non-Metricity in Gravity Theories}}.
\newblock PhD thesis, 2019.
\newblock
\href{http://www.arXiv.org/abs/1902.09643}{{1902.09643}}.
\newblock
% .

\bibitem{Iosifidis:2020gth}
D.~Iosifidis, \emph{{Cosmological Hyperfluids, Torsion and Non-metricity}},
  Eur. Phys. J. C {\bf 80} (2020), no.~11, 1042,
\href{http://www.arXiv.org/abs/2003.07384}{{2003.07384}}
% .

\bibitem{PhysRevLett.10.66}
R.~Penrose, \emph{Asymptotic Properties of Fields and Space-Times}, Phys. Rev.
  Lett. {\bf 10} (Jan, 1963)
66--68
% .

\bibitem{Penrose:1965am}
R.~Penrose, \emph{{Zero rest mass fields including gravitation: Asymptotic
  behavior}}, Proc. Roy. Soc. Lond. A {\bf 284} (1965)
159
% .

%\bibitem{Note1}
%Eq. \protect \eqref {eq:weyl} means that ${\protect \text {\protect \bf g}}$
%  and ${\protect \bm {\upupsilon }}$ are \protect \textit {Weyl-covariant
%  objects} of respective weights $-2$ and $1$.

%\bibitem{Note2}
%So far, we have only imposed the Einstein equations that provide, in Ashtekar's
%  terms~\cite {Ashtekar:1981hw,Ashtekar:1981bq}, the ``kinematic arena'' for
%  formulating the holographic correspondence, \protect \textit {i.e.}, the
%  nature of the boundary (conformal Carroll) and the connection induced on the
%  conformal boundary by the bulk physical Levi--Civita connection. These data
%  must be specified prior to formulating any holographic theory that is dual to
%  Einstein gravity. In AdS, for instance, one typically imposes, consistently
%  with the Einstein equations, that the conformal boundary is Lorentzian, with
%  an effective speed of light proportional to the cosmological constant, and
%  that the boundary connection is the Weyl--Levi--Civita connection.

\bibitem{Penrose_Rindler_1986}
R.~Penrose and W.~Rindler, {\em Conformal infinity}, p.~291–439.
\newblock Cambridge Monographs on Mathematical Physics.
\newblock Cambridge University Press,
1986.
\newblock
% .

\bibitem{Weyl:1918ib}
H.~Weyl, \emph{{Gravitation and electricity}}, Sitzungsber. Preuss. Akad. Wiss.
  Berlin (Math. Phys. ) {\bf 1918} (1918)
465
% .

\bibitem{10.4310/jdg/1214429379}
G.~B. Folland, \emph{{Weyl manifolds}}, Journal of Differential Geometry {\bf
  4} (1970), no.~2,
145 -- 153
% .

\bibitem{10.1063/1.529582}
G.~S. Hall, \emph{Weyl manifolds and connections}, Journal of Mathematical
  Physics {\bf 33} (07, 1992)
2633--2638
% .

\bibitem{Mars:1993mj}
M.~Mars and J.~M.~M. Senovilla, \emph{{Geometry of general hypersurfaces in
  space-time: Junction conditions}}, Class. Quant. Grav. {\bf 10} (1993)
  1865--1897,
\href{http://www.arXiv.org/abs/gr-qc/0201054}{{gr-qc/0201054}}
% .

\bibitem{Chandrasekaran:2018aop}
V.~Chandrasekaran, E.~E. Flanagan  and K.~Prabhu, \emph{{Symmetries and charges
  of general relativity at null boundaries}}, JHEP {\bf 11} (2018) 125,
\href{http://www.arXiv.org/abs/1807.11499}{{1807.11499}}
% .

\bibitem{Chandrasekaran:2021hxc}
V.~Chandrasekaran, E.~E. Flanagan, I.~Shehzad  and A.~J. Speranza,
  \emph{{Brown-York charges at null boundaries}},
\href{http://www.arXiv.org/abs/2109.11567}{{2109.11567}}
% .

\bibitem{Freidel:2022vjq}
L.~Freidel and P.~Jai-akson, \emph{{Carrollian hydrodynamics and symplectic
  structure on stretched horizons}}, JHEP {\bf 05} (2024) 135,
\href{http://www.arXiv.org/abs/2211.06415}{{2211.06415}}
% .

\bibitem{Ciambelli:2023mir}
L.~Ciambelli, L.~Freidel  and R.~G. Leigh, \emph{{Null Raychaudhuri: canonical
  structure and the dressing time}}, JHEP {\bf 01} (2024) 166,
\href{http://www.arXiv.org/abs/2309.03932}{{2309.03932}}
% .

\bibitem{Petkou:2022bmz}
A.~C. Petkou, P.~M. Petropoulos, D.~R. Betancour  and K.~Siampos,
  \emph{{Relativistic fluids, hydrodynamic frames and their Galilean versus
  Carrollian avatars}}, JHEP {\bf 09} (2022) 162,
\href{http://www.arXiv.org/abs/2205.09142}{{2205.09142}}
% .

\bibitem{Ciambelli:2018ojf}
L.~Ciambelli and C.~Marteau, \emph{{Carrollian conservation laws and Ricci-flat
  gravity}}, Class. Quant. Grav. {\bf 36} (2019), no.~8, 085004,
\href{http://www.arXiv.org/abs/1810.11037}{{1810.11037}}
% .

%\bibitem{Note3}
%Note that this is analogous to the standard treatment of matter fields coupled
%  to a background Lorentzian geometry, where the covariant (or
%  Einstein--Hilbert) energy--momentum tensor is obtained by varying the matter
%  action with respect to the background metric. Its conservation then follows
%  from diffeomorphism invariance and the matter equations of motion. The only
%  difference here is that the connection includes components not determined by
%  the geometry, thereby increasing the number of background fields.

\bibitem{deBoer:2017ing}
J.~de~Boer, J.~Hartong, N.~A. Obers, W.~Sybesma  and S.~Vandoren,
  \emph{{Perfect Fluids}}, SciPost Phys. {\bf 5} (2018), no.~1, 003,
\href{http://www.arXiv.org/abs/1710.04708}{{1710.04708}}
% .

\bibitem{deBoer:2021jej}
J.~de~Boer, J.~Hartong, N.~A. Obers, W.~Sybesma  and S.~Vandoren,
  \emph{{Carroll Symmetry, Dark Energy and Inflation}}, Front. in Phys. {\bf
  10} (2022) 810405,
\href{http://www.arXiv.org/abs/2110.02319}{{2110.02319}}
% .

\bibitem{deBoer:2023fnj}
J.~de~Boer, J.~Hartong, N.~A. Obers, W.~Sybesma  and S.~Vandoren,
  \emph{{Carroll stories}}, JHEP {\bf 09} (2023) 148,
\href{http://www.arXiv.org/abs/2307.06827}{{2307.06827}}
% .

%\bibitem{Note4}
%Note that flux-balance equations sharing the same structure as Eqs. \protect
%  \eqref {eq:flux balance general} appear while considering the constraints
%  implied by diffeomorphism invariance for relativistic theories coupled with
%  more general connections, \protect \textit {i.e.}, without requiring metric
%  compatibility or a vanishing torsion, see \cite
%  {Iosifidis:2019dua,Iosifidis:2020gth} and more recently \cite
%  {Iosifidis:2025sjx}.

%\bibitem{Note5}
%Performing a thorough analysis, one can check that the right-hand sides of Eqs.
%  \protect \eqref {eq:flux balance general} is a divergence if and only if the
%  hypermomentum $\protect \bm {\Omega }_0{}^a$ associated, in particular, with
%  the ``ambiguous'' part of the boundary connection vanishes. As we shall see
%  below, the latter is related to the news tensor which encodes the
%  gravitational radiation: sending it to zero would therefore strongly reduce
%  the bulk solution space.
%
\bibitem{Duval:1976ht}
C.~Duval and H.~P. Kunzle, \emph{{Dynamics of Continua and Particles from
  General Covariance of Newtonian Gravitation Theory}}, Rept. Math. Phys. {\bf
  13} (1978)
351
% .

%\bibitem{Note6}
%Namely, if $r$ is affine, then $\beta = 0$ everywhere and our gauge conditions
%  are those fixing Newman--Unti coordinates \cite
%  {Newman:1962cia,Barnich:2011ty,Ciambelli:2018wre}. Furthermore, if $r$ is
%  Sachs' luminosity distance (\protect \textit {i.e.}, the radius of expanding
%  null spheres), then our gauge fixing leads to Bondi--Sachs coordinates \cite
%  {Bondi:1962px,Sachs:1962wk}.

%\bibitem{Note7}
%This is a usual mechanism in Carroll physics, see \protect \textit {e.g.} \cite
%  {Henneaux:2021yzg} for a detailed explanation.

%\bibitem{Note8}
%By this, we mean a transformation $\protect \text {\protect \bf k}\DOTSB
%  \mapstochar \rightarrow \protect \text e^W \protect \text {\protect \bf k}$
%  where the function $W$ vanishes at the boundary.

\bibitem{Barnich:2011ty}
G.~Barnich and P.-H. Lambert, \emph{{A Note on the Newman-Unti group and the
  BMS charge algebra in terms of Newman-Penrose coefficients}}, Adv. Math.
  Phys. {\bf 2012} (2012) 197385,
\href{http://www.arXiv.org/abs/1102.0589}{{1102.0589}}
% .

\bibitem{Barnich:2012nkq}
G.~Barnich and P.-H. Lambert, \emph{{Asymptotic symmetries at null infinity and
  local conformal properties of spin coefficients}}, TSPU Bulletin {\bf 2012}
  (2012), no.~13, 28--31,
\href{http://www.arXiv.org/abs/1301.5754}{{1301.5754}}
% .

\bibitem{Compere:2018ylh}
G.~Comp\`ere, A.~Fiorucci  and R.~Ruzziconi, \emph{{Superboost transitions,
  refraction memory and super-Lorentz charge algebra}}, JHEP {\bf 11} (2018)
  200,
\href{http://www.arXiv.org/abs/1810.00377}{{1810.00377}}
% .

%\bibitem{Note9}
%Eq. \protect \eqref {eq:delta Cab} reproduces exactly Eq. (4.56) of \cite
%  {Barnich:2010eb}, up to three amendements: the sign convention in the
%  variation, the fact that $l = 0$, as it corresponds to $\theta $ here.

\bibitem{Henningson:1998gx}
M.~Henningson and K.~Skenderis, \emph{{The Holographic Weyl anomaly}}, JHEP
  {\bf 07} (1998) 023,
\href{http://www.arXiv.org/abs/hep-th/9806087}{{hep-th/9806087}}
% .

\bibitem{Balasubramanian:1999re}
V.~Balasubramanian and P.~Kraus, \emph{{A Stress tensor for Anti-de Sitter
  gravity}}, Commun. Math. Phys. {\bf 208} (1999) 413--428,
\href{http://www.arXiv.org/abs/hep-th/9902121}{{hep-th/9902121}}
%%CITATION = HEP-TH/9902121;%%.

\bibitem{Skenderis:2000in}
K.~Skenderis, \emph{{Asymptotically Anti-de Sitter space-times and their stress
  energy tensor}}, Int. J. Mod. Phys. A {\bf 16} (2001) 740--749,
\href{http://www.arXiv.org/abs/hep-th/0010138}{{hep-th/0010138}}
% .

\bibitem{Compere:2020lrt}
G.~Comp\`ere, A.~Fiorucci  and R.~Ruzziconi, \emph{{The $\Lambda$-BMS$_4$
  charge algebra}}, JHEP {\bf 10} (2020) 205,
\href{http://www.arXiv.org/abs/2004.10769}{{2004.10769}}
% .

\bibitem{Freidel:2021cjp}
L.~Freidel, R.~Oliveri, D.~Pranzetti  and S.~Speziale, \emph{{Extended corner
  symmetry, charge bracket and Einstein\textquoteright{}s equations}}, JHEP
  {\bf 09} (2021) 083,
\href{http://www.arXiv.org/abs/2104.12881}{{2104.12881}}
% .

\bibitem{Kmec:2024nmu}
A.~Kmec, L.~Mason, R.~Ruzziconi  and A.~Yelleshpur~Srikant, \emph{{Celestial
  Lw$_{1+\infty}$ charges from a twistor action}}, JHEP {\bf 10} (2024) 250,
\href{http://www.arXiv.org/abs/2407.04028}{{2407.04028}}
% .

%\bibitem{Note10}
%See also \cite {Rignon-Bret:2024gcx,Rignon-Bret:2024mef} for another proposal
%  for the covariant BMS charges derived from a Wald--Zoupas prescription \cite
%  {Wald:1999wa} and based on Geroch's supermomentum \cite {Geroch:1977big}.

\bibitem{Iosifidis:2025sjx}
D.~Iosifidis, M.~Karydas, A.~Petkou  and K.~Siampos, \emph{{On the geometric
  origin of the energy-momentum tensor improvement terms}},
\href{http://www.arXiv.org/abs/2503.21609}{{2503.21609}}
% .

\bibitem{Newman:1962cia}
E.~T. Newman and T.~W.~J. Unti, \emph{{Behavior of Asymptotically Flat Empty
  Spaces}}, J. Math. Phys. {\bf 3} (1962), no.~5,
891
% .

\bibitem{Henneaux:2021yzg}
M.~Henneaux and P.~Salgado-Rebolledo, \emph{{Carroll contractions of
  Lorentz-invariant theories}}, JHEP {\bf 11} (2021) 180,
\href{http://www.arXiv.org/abs/2109.06708}{{2109.06708}}
% .

\bibitem{Rignon-Bret:2024gcx}
A.~Rignon-Bret and S.~Speziale, \emph{{Centerless BMS charge algebra}}, Phys.
  Rev. D {\bf 110} (2024), no.~4, 044050,
\href{http://www.arXiv.org/abs/2405.01526}{{2405.01526}}
% .

\bibitem{Rignon-Bret:2024mef}
A.~Rignon-Bret and S.~Speziale, \emph{{Spatially local energy density of
  gravitational waves}}, JHEP {\bf 03} (2025) 048,
\href{http://www.arXiv.org/abs/2405.08808}{{2405.08808}}
% .

\end{thebibliography}

\providecommand{\href}[2]{#2}\begingroup\raggedright\endgroup

\end{document}